\documentclass[a4paper,twocolumn]{esapub} 

\usepackage[varg]{txfonts}
\usepackage{graphicx}
\usepackage{natbib}

\title{On the Mechanism of Type Ia Supernovae}
\author[(1)]{F. K. R{\"o}pke}
\author[(1)]{W. Hillebrandt}
\affil[(1)]{Max-Planck-Institut f\"ur Astrophysik,
  Karl-Schwarzschild-Str.~1, 85741 Garching,
Germany}
\author[(1),(2)]{S. I. Blinnikov}
\affil[(2)]{ITEP 117218, Moscow, Russia}

\newcommand{\aap}{A{\&}A}

\newcommand{\aj}{AJ}
\newcommand{\apj}{ApJ}
\newcommand{\apjl}{ApJL}

\newcommand{\mnras}{MNRAS}

\newcommand{\apss}{ApSS}

\begin{document}


\maketitle

\begin{abstract}
Type Ia supernovae (SNe Ia) are one of the major tools to determine the
cosmological parameters. Utilizing them as distance indicators, it is
possible to geometrically survey the universe. To this end, the
intrinsic scatter in the luminosities of these events needs to be
calibrated using empirical relations between observables. A
theoretical explanation for these relations is still lacking and can
only be provided by a sound understanding of the mechanism of SNe Ia. 
Recently there has been significant progress in modeling SNe Ia. We
report on numerical simulations of the explosion
process, compare the results with observations of nearby SNe
Ia, and discuss current uncertainties of the models. The presented
simulations shed some light on the origin of the diversity of SNe
Ia. Such simulations will pave the way towards an understanding of SN
Ia diversities and correlations of their properties and ultimately
provide a tool to validate the cosmological implications of SN Ia
distance measurements.
\end{abstract}

\section{Introduction}

Almost 80 years after Einstein's attempts to derive a cosmological
model from his theory of General Relativity, subsequently introducing
and discarding the cosmological constant, two cosmological surveys
\citep{riess1998a, perlmutter1999a}
of the universe based on distance measurements of type Ia supernovae
(SNe Ia) revolutionized the cosmological paradigm. They found that the
Universe is currently in a phase of \emph{accelerated} expansion. This
revived the discussion of the cosmological constant. Independent
determinations of cosmological parameters based on anisotropies in the
Cosmic Microwave Background radiations and on large-scale galaxy
surveys provided independent confirmation of the SN Ia measurements.

Yet the question of the applicability of SNe Ia as distance indicators
is still not satisfactory answered. SNe Ia are remarkably uniform
events by astrophysical standards,but evidently no
standard candles. Only the calibration of the distance measurements
according to empirical correlations between observables provides the
necessary accuracy for the determination of cosmological parameters. 
A firm theoretical reasoning of such correlations is, however, still
lacking.

It is still an open question whether the accelerated expansion of the
Universe is caused by a simple cosmological constant or some more
complicated constituent. Therefore one of the most fascinating
problems of cosmology is to determine the equation of state of this
``Dark Energy'' component. SNe Ia may be a suitable tool for this
task, but the accuracy of measurements required here can only be
reached if systematical errors that may arise from calibration of the
distance measurements are under control. A sound understanding of the
physical mechanism of SNe Ia is inevitable in this project.

This poses a great challenge to SNe Ia theory since it requires
self-consistent parameter-free modeling so that the predictive power of
the models is sufficient to tackle questions of the origin of
diversities and correlations.

Consequently, as a first step, an astrophysical model has to be built
and implemented into a numerical simulation code. As will be explained
below, the nature of relevant physical effects requires
three-dimensional simulations. This approach is notably
different from earlier one-dimensional SN Ia models which parametrized
physical effects and therefore could be tuned to fit the observational
data. Although this approach provided valuable hints to underlying
physical effects, it cannot be used to constrain the correlations.
Three-dimensional models have been implemented
over the past years by different groups on the basis of different
assumptions \citep[e.g.][]{hillebrandt2000b,reinecke2002d,gamezo2003a,
  calder2004a}. Most of them led to viable explosions and the task is now
to compare the results of these simulations with observations of
nearby SNe Ia. Once a model matching the observational requirements
has been established, it can be used to explore the diversity arising
from variations of physical parameters as suggested by progenitor
evolution and environmental effects. The results of such a survey should then be
analyzed to identify correlations in the characteristics. This may
eventually provide a theoretical basis for distance measurement
calibrations and a handle on possible systematical errors in those.

\section{Astrophysical model}

\subsection{Progenitor evolution and ignition}

The cornerstones of the astrophysical model of SNe Ia are set by two
fundamental characteristics of these events. Evidently, SNe Ia belong
to the most energetic cosmic explosions. For a short period of time
they can outshine an entire galaxy consisting of tens of billions of
stars. Assuming SNe Ia originate form single stellar objects, only
their gravitational binding energy, released in a collapse towards a
compact object, or its nuclear energy, released in explosive
reactions, come into consideration as possible energy sources. In the
particular case of SNe Ia no compact object is found in the remnant
excluding the first possibility. The fact that no hydrogen is found in
the spectra of SNe Ia provides a strong hint that the object
undergoing the nuclear explosion may be a white dwarf (WD) star consisting
of carbon and oxygen. 

A simple WD, however, is an inert object and thus the only way to
introduce the necessary dynamics into the system is to assume it to
be part of a binary system. Here the WD accretes matter -- in the
favored scenario from a non-degenerate companion -- until it approaches
the Chandrasekhar mass (the maximum mass that is supported against
gravitational collapse by the pressure of the degenerate electrons). At
this point, the density at the center of the WD increases rapidly so
that fusion of carbon ignites. This gives rise to a stage of
convective carbon burning lasting for several hundred years. This
phase is terminated by one or more small spatial regions undergoing a 
thermonuclear runaway, marking the birth of a thermonuclear flame
and the onset of the explosion. 

The limitation the fuel available in the explosion to the Chandrasekhar mass
($\sim$$1.4\,M_\odot$) makes the described scenario particularly
favorable since it provides a natural explanation for the striking
uniformity of SNe Ia in the gross observational features. On the other
hand, it is afflicted with great uncertainties. Achieving a stable
mass transfer in the progenitor binary system to build up a
Chandrasekhar mass WD is highly non-trivial \citep[e.g.][]{nomoto1985a} and the observational
evidence for such systems is sparse. The convective burning stage and
the conditions at flame ignition are extremely hard to model both
analytically and numerically. Therefore the exact shape and
configuration of the first flame spark(s) is not yet well constrained
\citep[e.g.][]{woosley2004a, hoeflich2002a}. 

\subsection{Flame propagation and explosion}
\label{sect:flame_explosion}

The goal of SN Ia explosion models is to follow the propagation of the
thermonuclear flame from its ignition near the center of the WD
outwards and to determine the composition and the distribution of the
burning products in the ejected material. 

In principle, hydrodynamics allows for two different modes of flame
propagation. One is the subsonic deflagration in which the flame is
mediated by the thermal conduction of the degenerate electron gas and
the other is a supersonic detonation in which the flame is driven by
shock waves. Already the first attempts to simulate SNe Ia \citep{arnett1969a}
revealed that a prompt detonation cannot explain these events since
the full star would be incinerated with sound speed. All material
would then be burnt at high densitites and converted to iron group
elements. Observed spectra of SNe Ia, however, clearly show indication
for significant amounts of intermediate mass elements (like Si, Ca, S
\ldots) in the ejecta. These can only be produced in the burning at
lower densities and therefore the material needs to expand prior to
incineration. This is provided if the flame propagation starts out as
a slow deflagration. A laminar deflagration front, however, propagates
too slowly to release sufficient energy to explode the WD star.

Thus, a key feature of a viable SN Ia model is the way of flame
acceleration. For this, two effects are concievable.

Firstly, the flame propagation from the center of the star outwards is
buoyancy unstable, since it leaves behind light and hot ashes below
the dense fuel -- a density statification inverse to the gravitational
acceleration. In its non-linear stage, the Rayleigh-Taylor instability
leads to the formation of mushroom-shaped burning bubbles raising into
the fuel. The Reynolds number typical for this situation is as high as
$10^{14}$. Clearly, shear (Kelvin-Helmholtz) instabilities at the
interfaces of these bubbles will generate turbulent eddies which then
decay to smaller scales foming a turbulent energy cascade. The flame
will interact with these eddies down to the Gibson-scale at which the
turbulent velocity fluctuations become comparable to the laminar flame
speed. This interaction corrugates the flame increasing its surface
and consequently enhances the net burning rate. Thus, the flame is
significantly accelerated and three-dimensional simulations
\citep[e.g.][]{hillebrandt2000b,reinecke2002d,gamezo2003a}
show that this effect may indeed lead to an explosion of the WD. In
this \emph{deflagration model}, SN Ia explosions are a problem of
turbulent combustion.

Secondly, a transition of the flame propagation mode from subsonic
deflagration to supersonic detonation could provide an additional
enhancement of the burning.  However, a mechanism
providing such a transition in a SN Ia explosion could not be found
yet and is therefore hypothetical. It can only be introduced at hoc
into a SN Ia model and is therefore free parameter -- an
effect that one would rather avoid in order to maintain the predictive
power of the model. 

We will thus focus on the pure deflagration model in the following.
After outlining the numerical implementation in Sect.~\ref{num_sect},
we will thus discuss the results from simulations and the
question to which degree they meet observable constraints in
Sect.~\ref{simobs_sect}.

\section{Numerical implementation}
\label{num_sect}

\subsection{Relevant scales}

The numerical implementation of a deflagration SN Ia model is
significantly complicated by the wide range of relevant length scales
involved in the problem. From the radius of the WD star
($\sim$$2000\,\mathrm{km}$ at the onset of the explosion and expanding
in the process) it reaches
down to the flame width and the Kolmogorov scale at which the
turbulent energy is dissipated -- both well below one
centimeter. Simulations capturing the entire star reach resolutions
around one kilometer while the Gibson scale is of the order of
$10^4\,\mathrm{cm}$ at the beginning of the explosion and decreases
steadily.

For large-scale SN Ia simulations this has three consequences.
\begin{enumerate}
\item \label{enum1} It is not possible to fully resolve the interaction of the flame
  with turbulence. Therefore modeling of the effects on unresolved
  scales is necessary.
\item \label{enum2} The internal flame structure cannot be
  resolved. Complementary small scale
  simulations in one \citep{timmes1992a} or more \citep{bell2004b} dimensions are required
  here. 
\item Assumptions about the flame properties at unresolved scales
  (e.g.\ stability below the Gibson scale) have to be validated in
  separate small-scale simulations \citep{roepke2003a,roepke2004a,roepke2004b}. 
\end{enumerate}

\subsection{Numerical techniques}

As obvious from the consideration of the relevant scales, a suitable
approach to implement a SNe Ia deflagration model on scales of the WD
is that of a Large Eddy Simulation (LES). It captures the effects of
turbulent phenomena at the largest scales where energy is fed into the
turbulent cascade and models turbulence on unresolved scales. The
numerical techniques forming the foundation of the implementation of a
deflagration SN Ia model described here are described in
\citet{reinecke1999a}, \citet{roepke2005c}, \citet{schmidt2005b}, and
\citet{niemeyer1995b}.

The hydrodynamics on the resolved scales is modeled in a finite volume
approach. It is based on the \textsc{Prometheus} implementation \citep{fryxell1989a}
of the piecewise parabolic method \citep{colella1984a}.

As a consequence of \ref{enum1} of the enumeration above, a
\emph{subgrid-scale model} is applied to account for turbulent effects on
unresolved scales. Some older simulations follow the implementation
suggested by \citet{niemeyer1995b}, while one recent highly resolved run applied the
updated modeling approach of \citet{schmidt2005b}.

According to \ref{enum2} from above, the flame representation has to
follow a modeling approach. Seen from the scales of the WD star, it
appears as a sharp discontinuity separating the fuel from the
ashes. A suitable numerical method to follow the evolution of such an
interface is the \emph{level set technique} introduced to SN Ia
modeling by \citet{reinecke1999b}.

In such an implementation, the flame propagation speed needs to be
prescribed. This is, however, not an arbitrary quantity in our
modeling framework. 
For turbulent combustion in the flamelet regime, which
applies to the burning in major parts of SN Ia explosions, it is given
by the turbulent velocity fluctuations. These are determined by the
subgrid-scale turbulence model.

Nuclear reactions are implemented in the simplified approach proposed
by \citet{reinecke2002b}. The progenitor material is composed of $^{12}$C and
$^{16}$O. At high fuel densities nuclear burning
terminates in nuclear statistical equilibrium represented by a mixture
of $^{56}$Ni and $\alpha$-particles. Once the fuel density drops below
$5.25 \times 10^7 \, \mathrm{g} \, \mathrm{cm}^{-3}$ burning will become
incomplete and intermediate mass elements (represented by  $^{24}$Mg)
are produced. The respective differences in nuclear binding energy is
released which provides sufficient accuracy to model the dynamics of
the explosion. 

However, in order to derive synthetic spectra and light curves from models, the
accurate distribution of different chemical species in the ejecta
needs to be determined. This is achieved in a postprocessing step
which makes use of the temperature and density evolution recorded by
tracer particles advected in the hydrodynamical explosion simulation. On the
basis of this data, the details of the nuclear reactions in the ashes
are reconstructed employing an extended nuclear reaction network
\citep{travaglio2004a}. 

\section{Simulations vs. observations}
\label{simobs_sect}

\begin{figure*}[t]
\centering
\includegraphics[width = \linewidth, clip]{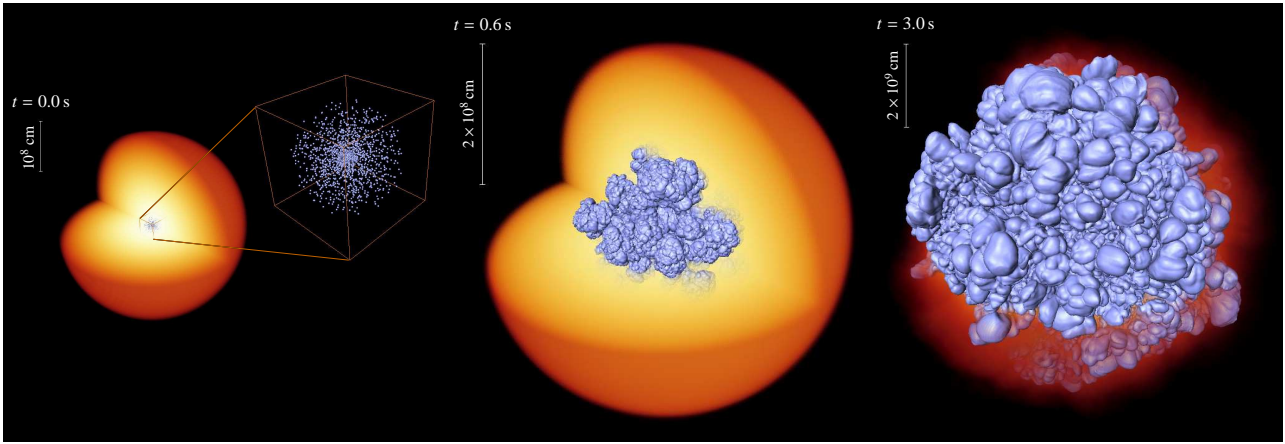}
\caption{Snapshots from a high-resolved SN Ia simulation starting from
  a multi-spot ignition scenario. The logarithm of the density is
  volume rendered indicating the extend of the WD star and the
  isosurface corresponds to the thermonuclear flame.\label{fig:evo}}
\end{figure*}

\subsection{Explosion models and global characteristics}

We will now review the status of deflagration SN Ia modeling with
particular emphasis on the question how the results of such
simulations compare with observations. Due to recent progress in such
simulations, a direct comparison with details of observations of nearby
SNe Ia has come into reach.

One requirement to reliably derive observables from simulations is that
the evolution of the models be followed to the stage of homologous
expansion. In this hydrodynamically relaxed situation, the velocity of
the ejected material is proportional to its radius. The implementation
of a moving computational grid on which the simulations are carried
out facilitated the long-term evolution of expanding SN Ia models
\citep{roepke2005c}. 

Due to the high computational expenses, most three-dimensional
simulations comprised only one spatial octant of the WD star, assuming
mirror symmetry to the other octants. However, only full-star setups
(such as presented by \citet{roepke2005b}) allow to study asymmetry effects.
On the basis of spectrapolarimetry observations of several SNe Ia
\citep[e.g.][]{wang2003a}
these are expected to occur in at least some explosions. 
\citet{roepke2005b} showed, that such
asymmetries arise exclusively from asymmetries in the flame ignition
conditions and not from large-scale instabilities and resulting
prefered modes in the flow patterns. Therefore, fixing a symmetric initial
flame shape and studying the influence of other physical parameters on
single-octant explosion models is a valid approach.

The propagation of the thermonuclear flame through the WD is shown in
Fig.~\ref{fig:evo}. The snapshots are taken from a recent
high-resolution simulation in which the flame was ignited in multiple
sparks around the center of the star (cf.\ the left part of
Fig.~\ref{fig:evo}). It agrees with the expectations
in the astrophysical scenario outlined in
Sect.~\ref{sect:flame_explosion}. After ignition, the Rayleigh-Taylor
instability leads to the formation of mushroom-shaped burning bubbles
on large scales which are subsequently superposed by smaller modes of
the instability. The resulting flow quickly becomes turbulent and the
flame is being corrugated. This is apparent in the second snapshot of
Fig.~\ref{fig:evo}. In this turbulent burning mode, the flame
propagates through most parts of the star (cf.\ the third snapshot in
Fig.~\ref{fig:evo}). A powerful explosion of the WD star results in
this simulation.

Naturally, the question arises if the outcome of simulations as the
one described above meets the observational constraints. Such
constraints arise from the global characteristics derived from
observations, observed light curves, and spectra taken from nearby SNe
Ia.

The global characteristics derived from SN Ia observations state that
a valid explosion model should release around $10^{51}\,\mathrm{erg}$
of energy and produce $\sim$$0.4\ldots 0.7\, M_\odot$ of $^{56}$Ni in
the nuclear burning. However, there exists a large diversity in the
observations ranging from the class of sub-luminous SNe Ia (like
SN~1991bg) to super-luminous events (e.g.\ SN~1991T). Our deflagration
models started with a multi-spot ignition setup typically possess
$7\ldots 8 \times 10^{50}\,\mathrm{erg}$ of asymptotic kinetic energy
of the ejecta. They produce around $0.4 \, M_\odot$ of
$^{56}$Ni. Thus they fall into the range of observational
expectations, but in the current stage do not account for the more
energetic SNe Ia.

\begin{figure}[t]
\centering
\includegraphics[width = \linewidth, clip]{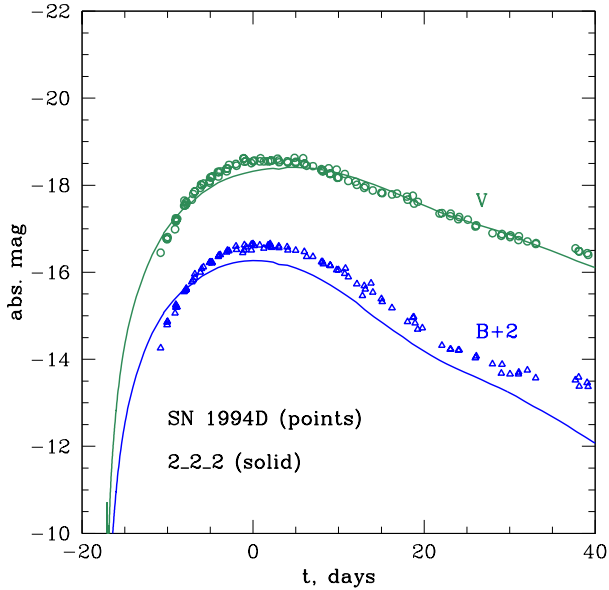}
\caption{Synthetic light curves derived from model 2\_2\_2 of
  \citep{roepke2005d} (solid curves) compared
  with observed light curves from SN~1994D.\label{fig:lc}}
\end{figure}

\subsection{Lightcurves}

A further requirement is that synthetic lightcurves agree with
observed ones. These are sensitive to the energy release, the
$^{56}$Ni production, as well as to the distribution of elements in
the ejecta. In Fig.~\ref{fig:lc} we compare synthetic light curves
derived from the \emph{2\_2\_2} simulation \citep{roepke2005d} with
observations of SN~1994D.
The multi-band light curve of this model was
calculated using the \textsc{Stella} code of \citet{blinnikov1998a} and
\citet{blinnikov2000a}.

The distance  to  SN~1994D is still controversial.
We have assumed the distance modulus 30.4 from \citet{drenkhahn1999a}

The model produced $0.3 \, M_\odot$ of $^{56}$Ni, although observations
require a bit higher $^{56}$Ni mass ($\sim 0.4 \, M_\odot$) for the assumed
distance. There is
generally very good agreement in the  B and V bands near
peak luminosities and in decline rate 20 days after the peak
which is most important for cosmological applications of type Ia supernovae.

\subsection{Spectra}

A much harder test for the models is posed by the comparison of
synthetic with observed spectra since these depend on details in the
compositon of the ejected material.

\begin{figure}[t]
\centering
\includegraphics[width = \linewidth, clip]{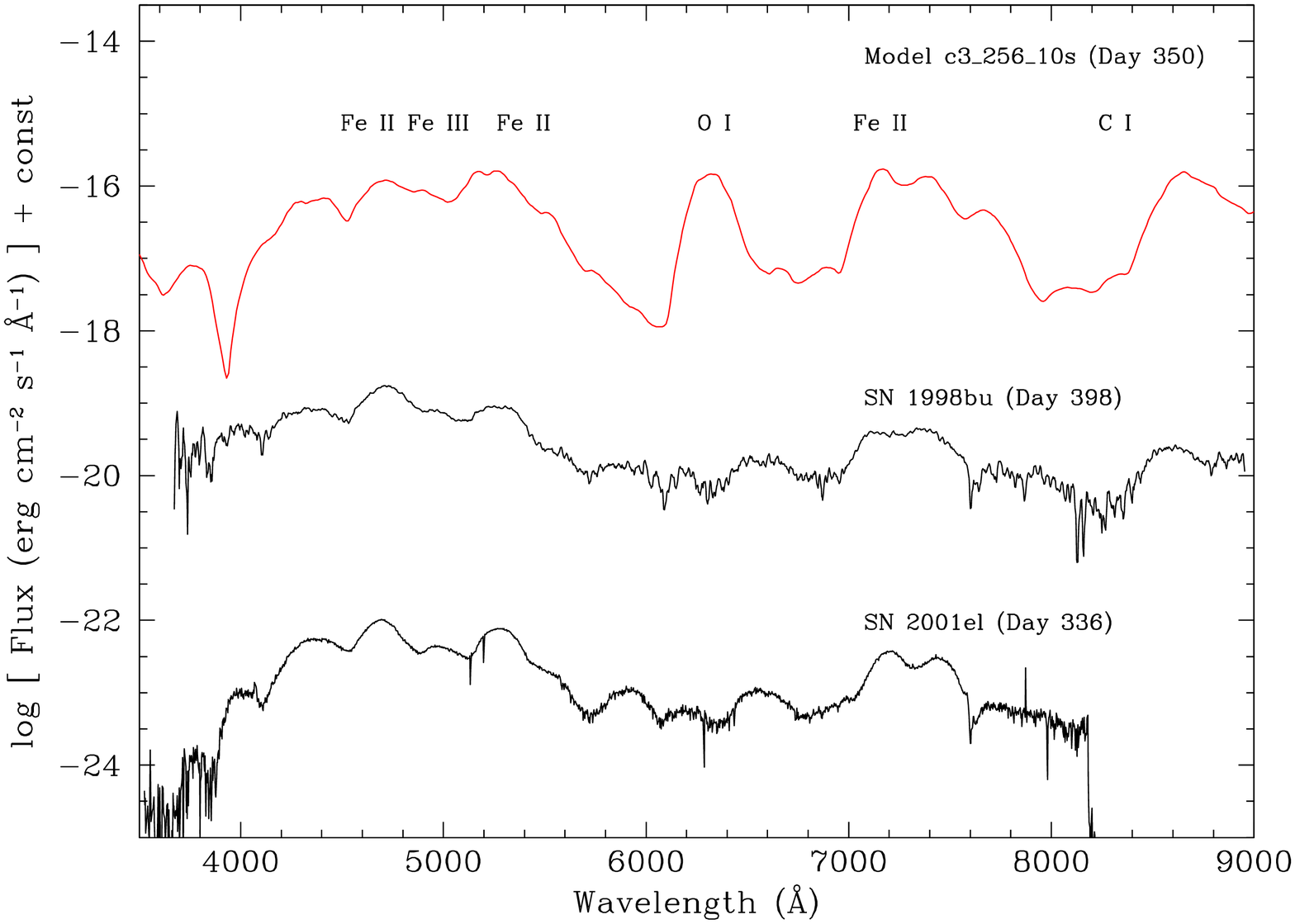}
\caption{Synthetic nebular spectrum compared with observations
  \citep[from][]{kozma2005a}.\label{fig:ns}}
\end{figure}

Nebular spectra provide a means of studying the central parts of the
ejecta, since they are taken at epochs where these have become
transparent due to expansion. Thus they explore the ``heart of the
explosion'' and are a valuable tool to study details of the physical
processes involved in the explosion stage. Unfortunately, only one
single synthetic late time spectrum is available from our deflagration
SN Ia models \citep{kozma2005a}. It was derived from a run that was intended to
test the implementation of the moving grid and is not believed to
provide a realistic SN Ia model. The flame was ignited in an
axisymmetric shape where a sphere around the center of the
WD star was perturbed by a toroidal structure of large wavelength and
amplitude. This artificial and simple initial flame shape
gives reason to not expect a good agreement between model and
observation. However, as shown in Fig.~\ref{fig:ns}, the broad iron
features of the observed spectra are qualitatively reproduced. An
inconsistency of the model with the observed nebular spectra is the
appearance of a pronounced oxygen line at $6300\,\mathrm{\AA}$. Both features of the synthetic
spectrum share a common origin. The broad iron lines are caused by NSE
material that is transported in the uprising plumes of ashes and thus
distributed in velocity space. At the same time strong downdrafts
carry unburnt material towards the center of the WD.

The disagreement may in part be attributed to the simplicity of the
explosion model. Its highly symmetric initial flame shape with large
imprinted perturbations favor a pronouned evolution of
large-scale Rayleigh-Taylor features. To answer the question whether
the unburnt material at low velocities is a generic problem in the
deflagration model of SNe Ia, the high-resolved simulations mentioned
above was carried out. The finer structure of the turbulent flame in
this simulation together with improved flame modeling led to a better
burnout of the downdrafts. Much less unburnt material is found at low
velocities in this model.    

A powerful diagnostic tool to compare SN Ia models with observations
is provided by the abundance tomography presented by
\citet{stehle2005a}. It makes 
use of spectra taken from SN~2002bo with an extraordinary good time
coverage. Fitting this sequence of data with synthetic spectra
unveils the composition of the ejecta in velocity space slice by
slice, since the photospere moves gradually inwards with the expansion
of the remnant. This abundance tomography of the ejecta can be
compared with results of our three-dimensional models, when averaged
over the angles. Qualitatively, the mixed composition of the ejecta
found by \citet{stehle2005a} is reproduced by deflagration SN Ia
models in a natural way since these predict a distribution of burnt
material with the rising bubbles. A problem was, however,
that older predicted large unburnt material fractions in the central
parts of the ejecta in disagreement with the results of
\citet{stehle2005a}. Our recent high-resolved simulation cures
this problem by clearly reproducing the iron-group dominance in the
low-velocity ejecta. The results are currently being analyzed and seem
promising for matching the observations.

\section{Diversity and correlations}

\begin{table}
  \begin{center}
    \caption{Variation of initial parameters in SN Ia explosion
      models.}\vspace{1em}
    \renewcommand{\arraystretch}{1.25}
    \begin{tabular}[h]{lrrr}
      \hline
      Parameter & \multicolumn{1}{p{0.155 \linewidth}}{range of\newline
      variation} & 
      \multicolumn{1}{p{0.17 \linewidth}}{effect on $^{56}$Ni
      production} & \multicolumn{1}{p{0.15 \linewidth}}{effect on
      total energy} \\
      \hline
      X($^{12}$C)  & [0.30,0.62]  & $\le$$2\%$ & $\sim$14\% \\
      $\rho_\mathrm{c}$ [$10^9 \, \mathrm{g}/\mathrm{cm}^3$]  &
      [1.0,2.6] & $\sim$6\% & $\sim$17\% \\
      $Z$ [$Z_\odot$] & [0.5,3.0] & $\sim$20\% & none \\
      \hline \\
      \end{tabular}
    \label{tab:res}
  \end{center}
\end{table}

The recent developments in the deflagration SN Ia explosion modeling
outlined in the previous section seem to indicate that such a model is
capable of reproducing the main features of observed objects. They do
not rule out the alternative of a delayed detonation. The success of
the pure deflagration model, however, may be taken as evidence that
this stage provides at least a major contribution to the explosion
process and general properties will not be changed much by a
hypothetical deflagration-to-detonation transition.

It is therefore an important question to ask how such a model behaves
under variation of physical parameters. Does it reproduce the observed
diversity of SNe Ia? Are correlations between observables evident in
the results?

Unfortunately, three-dimensional deflagration models of SNe Ia as
described above are computationally expensive. To moderate these
expenses, simplified setups may be used to study effects of physical
parameters on the explosion models. Such an approach was recently
taken by \citet{roepke2005d} and resulted in the first systematic study of
progenitor parameters in three-dimensional models. The basis of this
study was a single-octant setup with moderate (yet numerically
converged) resolution. However, the lack in resolution did not allow a
reasonable multi-spot ignition scenario and thus only weak explosions
can be expected. It is therefore not possible to set the absolute
scale of effects in this approach, but trends can clearly be
identified. 

The parameters chosen for the study were the WD's
carbon-to-oxygen ratio, its central density at ignition and its
$^{22}$Ne mass fraction resulting from the metallicity of the
progenitor. All parameters were varied independently to study the
individual effects on the explosion process. In a realistic scenario,
however, these parameters are interrelated by the evolution of the
progenitor binary system. The results of this survey are given in
Tab.~\ref{tab:res}.

A variation of the carbon-to-oxygen ratio effects the energy production
in the burning due to the differences in the binding energies of these
two nuclei. Counterintuitively, this results in no significant change
in the $^{56}$Ni production. \citet{roepke2004c} explained this effect by the fact
that the potentially higher energy release in carbon-rich models is
buffered by a higher $\alpha$-particle fraction in the NSE and only
released when burning is already incomplete.

Models with lower central densities show a delayed flame
evolution and consequently a lower and delayed energy production. This
is due to the fact that the flame experiences a lower gravitational
acceleration in these models resulting in decreased turbulence
generation. Therefore less $^{56}$Ni is produced in these models. This
effect is even more pronounced due to the fact that in low-density WDs less
material is present at sufficient densities to be potentially burnt to
NSE. A counteractive effect is expected at higher densities. Here,
electron captures will become important favoring neutron-rich isotopes
over $^{56}$Ni in the NSE. The dynamical effects of electron captures are,
however, not yet implemented in our explosion model and therefore the
current survey is restricted to relatively low central densities.

\begin{figure}[t!]
\centering
\includegraphics[width = \linewidth, clip]{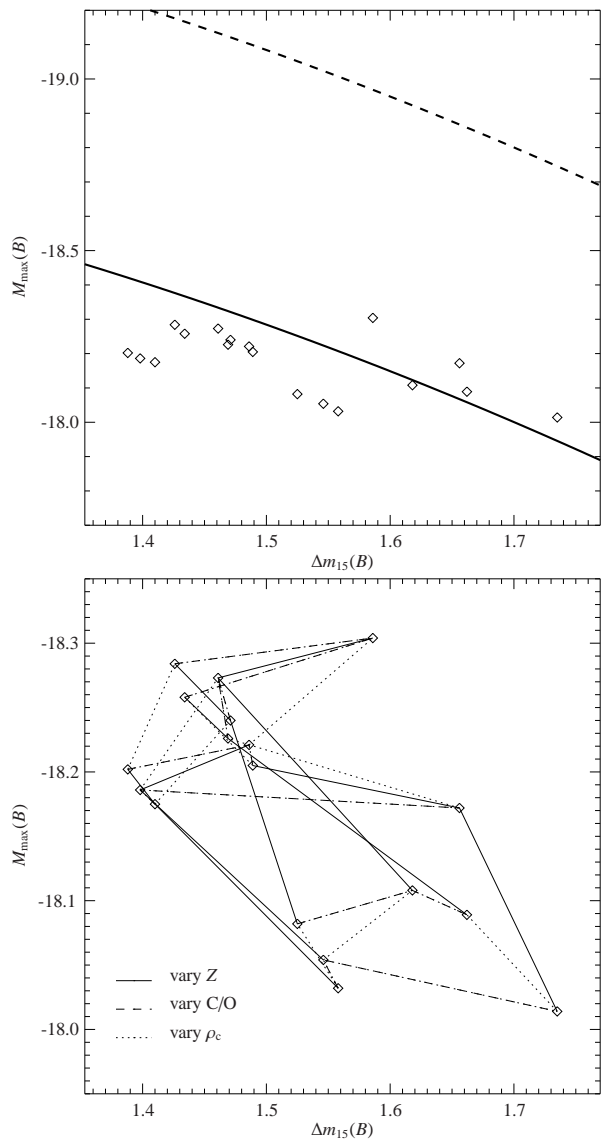}
\caption{Peak luminosity vs. decline rate of the light curve in the B
  band (diamonds correspond to SN Ia explosion models). Compared with
  original relation by \citet{phillips1999a}
  (dashed curve) and shifted relation (solid curve) in the top
  panel.\label{fig:ph}}
\end{figure}

The metallicity of the progenitor star results in a certain $^{22}$Ne
mass fraction in the WD. This is an isotope with neutron excess and
therefore again favors the production of neutron-rich species over
$^{56}$Ni in the NSE. Our results confirm the analytic prediction by
\citet{timmes2003a} and agree with \citet{travaglio2005a}. The
metallicity parameter, however, has no effect on the explosion
dynamics and the energy production in our models.

To determine the effects of these variations on observables, synthetic
light curves were derived from all models (an example is shown in
Fig.~\ref{fig:lc}). From these, the peak
luminosities and decline rates (in magnitudes 15 days after maximum;
$\Delta m_{15}$), were determined. The pioneering work by
\citet{phillips1993a}
established the relation between the peak luminosity and $\Delta
m_{15}$ as one of the primary tools to calibrate cosmological SN Ia
distance measurements. The so-called \emph{Phillips relation}
quantifies the decrease of $\Delta m_{15}$ for brighter SNe Ia.

The results from our models are compared with the relation given by
\citet{phillips1999a} in Fig.~\ref{fig:ph}. Obviously, the absolute magnitude of the
\citet{phillips1999a} relation is not met by our set of models (cf.\ the upper panel of
 Fig.~\ref{fig:ph}). Moreover, the range
of scatter in $\Delta m_{15}$ is much narrower than that of the set of
observations used by \citet{phillips1999a}. The trend of our models is not
clear. They are best fit by a line with a slope consistent with the
\citet{phillips1999a} relation. It is, however, obvious, that a better agreement
cannot be expected given the fact, that the parameters in our survey
were chosen independently. A consistent stellar evolution would pick a
sub-sample of our set of models and it will be investigated in forthcoming
studies whether this will be in agreement with the Phillips relation.

Nonetheless, with our set of models, the question can be answered,
which parameter dominates the slope in the direction of dimmer events
for faster decline rates. The varied parameters are coded by different
line styles in the lower panel of Fig.~\ref{fig:ph}. Clearly, the
progenitor's metallicity can be identified as this
parameter. Variations in the central density and the carbon mass
fraction of the WD superpose a scatter on the dominant relation.

\section{Conclusions}

We discussed the mechanism of SN Ia explosions and reviewed the common
astrophysical scenarios applied to explain these events. The
implementation of such models into numerical simulations was described
focusing on the pure deflagration model for SNe Ia. Subsequently,
recent progress in numerical simulations was reported. On the basis
of comparison of these results with observations, the conclusion is
drawn, that pure deflagration models are at least capable of
reproducing the gross features of observed SNe Ia.

Such models therefore can be applied to study the origin of the
diversity of SNe Ia. In a first systematic survey of initial
parameters of three-dimensional models, the effects of the WD's
carbon-to-oxygen ratio, its central density and the progenitor's
metallicity on the explosion process were analyzed. Although this was
only possible on the basis of simplified and therefore weakly
exploding models, clear trends could be identified.

From this set of models, synthetic light curves were derived. The
relation between the peak-luminosities and the decline rates of the
light curves were found to be consistent with the slope predicted by
\citet{phillips1999a}. Due to the simplicity of the applied models, the absolute
magnitude of effects could not be predicted.

Forthcoming studies will focus on the questions how a realistic
stellar evolution of the progenitor interrelates the effects. A second
important question is whether the range of scatter predicted by the
observations can be reproduced with more elaborate models. Although it
captures main aspects of observations, it may be possible that the
applied SN Ia model is still incomplete and additional physical
effects (like delayed detonations) need to be taken into account. 

\section*{Acknowledgments}

The work presented in this contribution is a result of intensive
collaboration with M.~Reinecke, E.~Sorokina, J.~Niemeyer, W.~Schmidt,
M.~Gieseler, C.~Travaglio, C.~Kozma, M.~Stehle and P.~Mazzali. We
would like to thank the organizers of the EPS 13 conference. FKR's
conference participation was kindly supported by the European Southern
Observatory.

\end{document}